\documentclass[prl,superscriptaddress,reprint,showpacs,longbibliography]{revtex4-2}
\usepackage{lineno}
\usepackage{color}
\usepackage{physics}
\usepackage{graphicx,textcomp,amssymb,amsmath,dcolumn,hyperref}
\usepackage{siunitx}
\usepackage{comment}

\begin{document}

\title{Pauli Blockade of Tunable Two-Electron Spin and Valley States in Graphene Quantum Dots}

\author{Chuyao Tong}
\email{ctong@phys.ethz.ch}
\author{Annika Kurzmann}
\author{Rebekka Garreis}
\author{Wei Wister Huang}
\author{Samuel Jele}
\author{Marius Eich}
\author{Lev Ginzburg}
\author{Christopher Mittag}
\affiliation{Solid State Physics Laboratory, ETH Zurich, CH-8093 Zurich, Switzerland}

\author{Kenji Watanabe}
\affiliation{Research Center for Functional Materials, National Institute for Materials Science, 1-1 Namiki, Tsukuba 305-0044, Japan}
\author{Takashi Taniguchi}
\affiliation{International Center for Materials Nanoarchitectonics, National Institute for Materials Science,  1-1 Namiki, Tsukuba 305-0044, Japan}

\author{Klaus Ensslin}
\author{Thomas Ihn}
\affiliation{Solid State Physics Laboratory, ETH Zurich, CH-8093 Zurich, Switzerland}

\date{\today}

\begin{abstract}
Pauli blockade mechanisms --- whereby carrier transport through quantum dots is blocked due to selection rules even when energetically allowed --- are a direct manifestation of the Pauli exclusion principle, as well as a key mechanism for manipulating and reading out spin qubits. Pauli spin blockade is well established for systems such as GaAs QDs, but is to be further explored for systems with additional degrees of freedom, such as the valley quantum numbers in carbon-based materials or silicon. Here we report experiments on coupled bilayer graphene double quantum dots, in which the spin and valley states are precisely controlled, enabling the observation of the two-electron combined blockade physics. We demonstrate that the doubly occupied single dot switches between two different ground states with gate and magnetic-field tuning, allowing for the switching of selection rules: with a spin-triplet--valley-singlet ground state, valley-blockade is observed; and with the spin-singlet--valley-triplet ground state, robust spin blockade is shown.

\end{abstract}
\maketitle
\begin{figure*}
	\includegraphics[width=17.8cm]{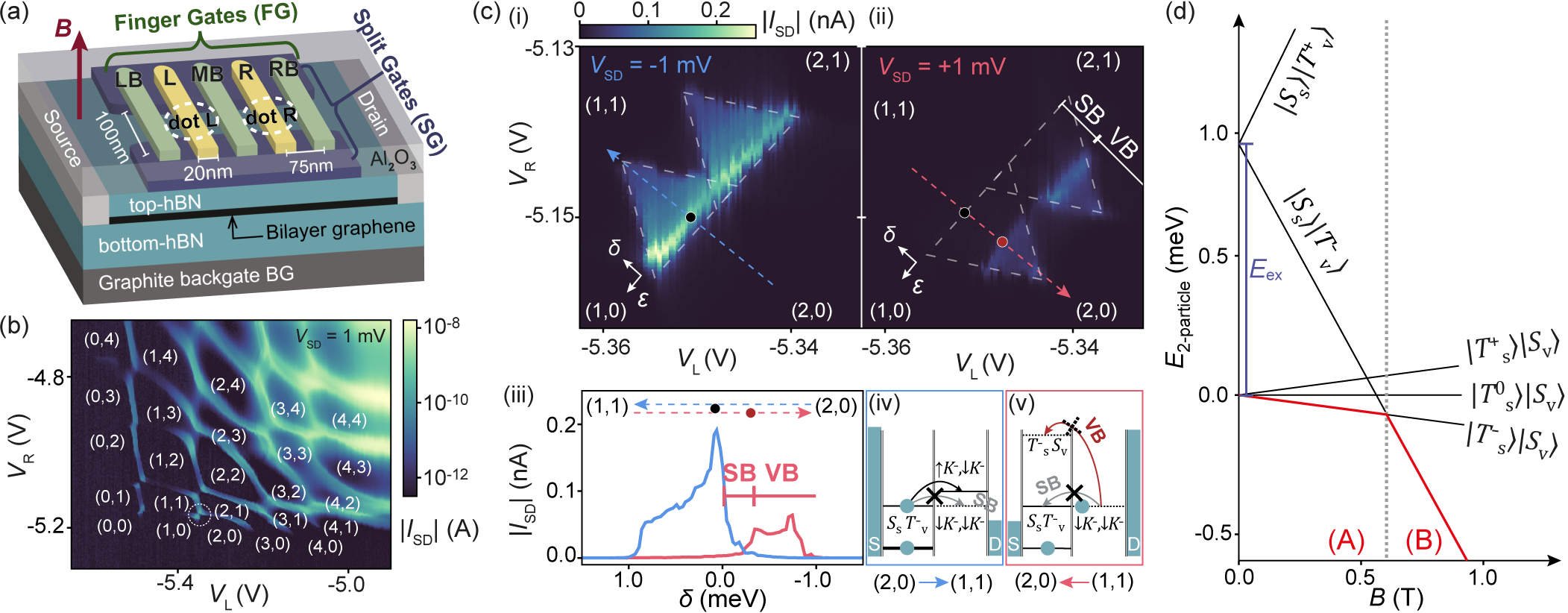}
	\caption{(a) Device illustration. Double electron dots are defined with plunger-gates L and R, and barrier-gates $\mathrm{LB}$, $\mathrm{MB}$ and $\mathrm{RB}$. A magnetic field $B$ is applied perpendicular to the BLG. (b) Charge-stability diagram at $V_\mathrm{B_M}=\SI{-5.76}{V}$. (c) Finite-bias triangles at $V_\mathrm{MB}=\SI{-5.76}{V}$, $B=\SI{800}{\milli T}$ with (i) negative, and (ii) positive source--drain bias $V_\mathrm{SD}$. $\epsilon$ is the total energy, and detuning $\delta=\epsilon_\mathrm{L}-\epsilon_\mathrm{R}$ the inter-dot energy difference. (iii) Line-cuts along the dashed arrows, with $V_\mathrm{L,R}$ converted into $\delta$. $\delta=0$ at the baseline of bias-triangles. Current peaks are labeled by dots. Valley-blockade (VB) and spin-blockade (SB) suppress current in the positive-bias direction. Energies of relevant states are sketched for $\delta=0$ in (iv) and (v). For negative-bias (electron transport $(2,0)\rightarrow(1,1)$), the spin-blockaded GS--GS transition (gray in (iv)) is readily circumvented by a transition close in energy (black). For positive-bias (electron transport $(1,1)\rightarrow(2,0)$), the next available transition is higher in energy (red in (v)) and requires a valley flip. (d) Evolution of single-dot two-particle energies in magnetic field, sketched with $E_\mathrm{ex}=\SI{0.9}{\milli eV}$, $g_\mathrm{v}=28$ and $g_\mathrm{s}=2$. The two different GSs (red) define regime A and B.}
	\label{figstart}
\end{figure*}

\begin{figure*}
	\includegraphics[width=17.8cm]{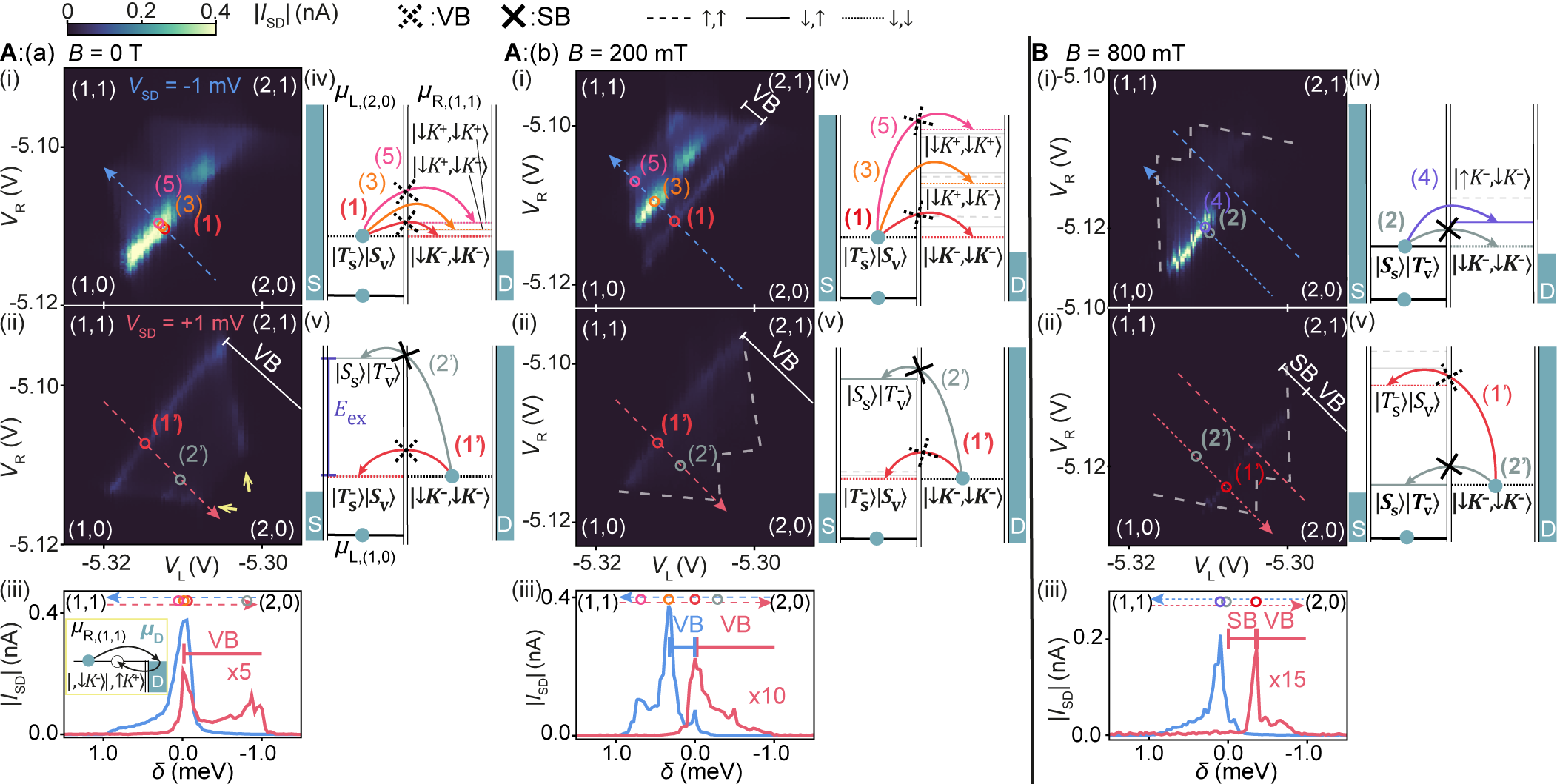}
	\caption{Finite-bias triangles at $V_\mathrm{MB}=\SI{-5.81}{V}$ in \textbf{A}: regime A with $(2,0)_\mathrm{GS}:\ket{T^-_\mathrm{s}}\ket{S_\mathrm{v}}$ at $B=$ (a) $\SI{0}{T}$ and (b) $\SI{200}{mT}$, and in \textbf{B}: regime B with $(2,0)_\mathrm{GS}:\ket{S_\mathrm{s}}\ket{T^-_\mathrm{v}}$ at $B=\SI{800}{mT}$ for (i) negative (electron transport $(2,0)\rightarrow(1,1)$), and (ii) positive (electron transport $(1,1)\rightarrow(2,0)$) source--drain bias $V_\mathrm{SD}$. (iv,v) Schematics of electrochemical potentials $\mu$ of relevant transitions for (i,ii), sketched at $\delta=0$, when $\mu_\mathrm{GS}$s align. Non-zero $\delta$ allows for higher energy transitions. (iii) Line-cuts along the dashed arrow, with $V_\mathrm{L,R}$ converted into $\delta$. Current resonances are labeled by numbered colored circles. Valley-blockade (VB) and spin-blockade (SB) regions are marked. Current is suppressed for positive bias $(1,1)\rightarrow(2,0)$ by the valley- and spin-blockade, and is enhanced for clarity in the line-cuts (iii) by a factor of 5, 10, and 15 for A(a), A(b), and B. In A(a,ii), the valley blockade is lifted at the edges of the triangles, where the electron in the dot can exchange with electrons in the source or drain leads (yellow arrows in A(a,ii) and schematic outlined in yellow in A(a,iii)). In B, the line-cut is taken at the dotted instead of the dashed line, due to a shift of the bias triangle along the $\epsilon$ direction, a result of the change of $(2,0)_\mathrm{GS}$.}
	\label{figtriangle}
\end{figure*}

Graphene quantum dots (QDs) have been proposed to host spin qubits with long spin coherence times \cite{Trauzettel2007spinqubit,RMPspin}, especially promising in bilayer graphene (BLG) due to its smaller spin--orbit coupling compared to that of carbon nanotubes \cite{pei2012valley,CNTnuclear2009,CNTblockadePRB,AnnikaKondo, Luca1001}. In BLG, a band-gap can be opened by an electric field perpendicular to the BLG sheet \cite{Ohta2006BLGband,mccann2006BLGband,Oostinga2008BLG}. Together with recent advancements in fabrication technology \cite{Hiske2018electrostatically}, the quality of state-of-the-art BLG QDs has been raised to such a level that highly tunable QDs \cite{MariusPRX, Tunabledot, RebekkaPRL,annikaexcitedstates,AnnikaKondo, Luca1001, Lucacrossover, LucaInterdot} can now be fabricated.

Observation of Pauli blockade is a crucial step towards qubit manipulation and read out. A coupled double QD occupied by two carriers can be tuned to a regime where two states coexist: one carrier on each dot, or both carriers on the same dot. Transitions between these states can be blocked by selection rules based on Pauli exclusion principle. Observation of two-electron Pauli spin blockade relies on the single-dot two-electron spin-singlet and -triplet states to be well-separated in energy, with the spin-singlet being the ground state (GS). This usually arises naturally at zero magnetic field \cite{ono2002current,petta2005coherent,JohnsonBlockade,CNTnuclear2009,CNTblockadePRB}. In BLG QDs however, at low magnetic field, the single-dot two-carrier GS is observed to be a spin-triplet \cite{annikaexcitedstates,RebekkaPRL,aachen2electron} due to the additional valley degrees of freedom $K^{-/+}$. 

Unlike the low-lying valleys in silicon, complicating qubit control by providing additional coherence channels \cite{RevModPhys.85.961,maune2012coherent, kawakami2014electrical}, energy splittings of BLG valley states are reliably tunable by perpendicular magnetic fields \cite{MariusPRX,annikaexcitedstates,RebekkaPRL, Tunabledot} and by gate voltages \cite{Tunabledot,chen2020gate}, and are themselves good quantum numbers. Selection rules involving valleys have been seen in carbon nanotubes and silicon \cite{perron2017valley,CNTnuclear2009,CNTblockadePRB}, although with limited control. In our coupled double QDs, valley tunability allows us to study the combined spin and valley blockade physics, demonstrating controlled switching between a spin-triplet--valley-singlet at low, and spin-singlet--valley-triplet single-dot two-electron state at high magnetic field. In this way, we show canonical two-electron blockade physics by performing finite-bias measurements, and observe valley blockade in the former and spin blockade in the latter regime.

We utilized the tunable BLG bandgap \cite{Ohta2006BLGband,mccann2006BLGband,Oostinga2008BLG} to form a coupled electron double QD with $n$-type leads [Fig.~\ref{figstart}(a), see Supplemental Materials S1]. Barrier-gate (green) voltages $V_\mathrm{LB,MB,RB}$ provide individual control of the dot-lead \cite{mariusnanolett,Tunabledot}, and inter-dot tunnel coupling \cite{LucaInterdot}. Dots L, R are independently controlled by the plunger-gate (yellow) voltages $V_\mathrm{L,R}$. A bias-voltage $V_\mathrm{SD}$ is applied symmetrically between source ($+V_\mathrm{SD}/2$) and drain ($-V_\mathrm{SD}/2$), and the current is measured. 

The charge-stability map [Fig.~\ref{figstart}(b)] displays honeycomb patterns. Regions of low conductance suppressed by Coulomb-blockade are labeled $(N_\mathrm{L},N_\mathrm{R})$, with stable electron numbers in the left $(N_\mathrm{L})$ and in the right $(N_\mathrm{R})$ dot. Transport resumes at intersections of Coulomb resonances of the two dots and pairs of triple-points of high conductance appear. Three double dot charge occupancies coexist at the triple-point that they are adjacent to, and allow for charge transport via these three states. More negative plunger-gate voltages $V_\mathrm{L,R}$ deplete the respective dots down to the last electron.

\begin{figure*}
	\includegraphics[width=17.8cm]{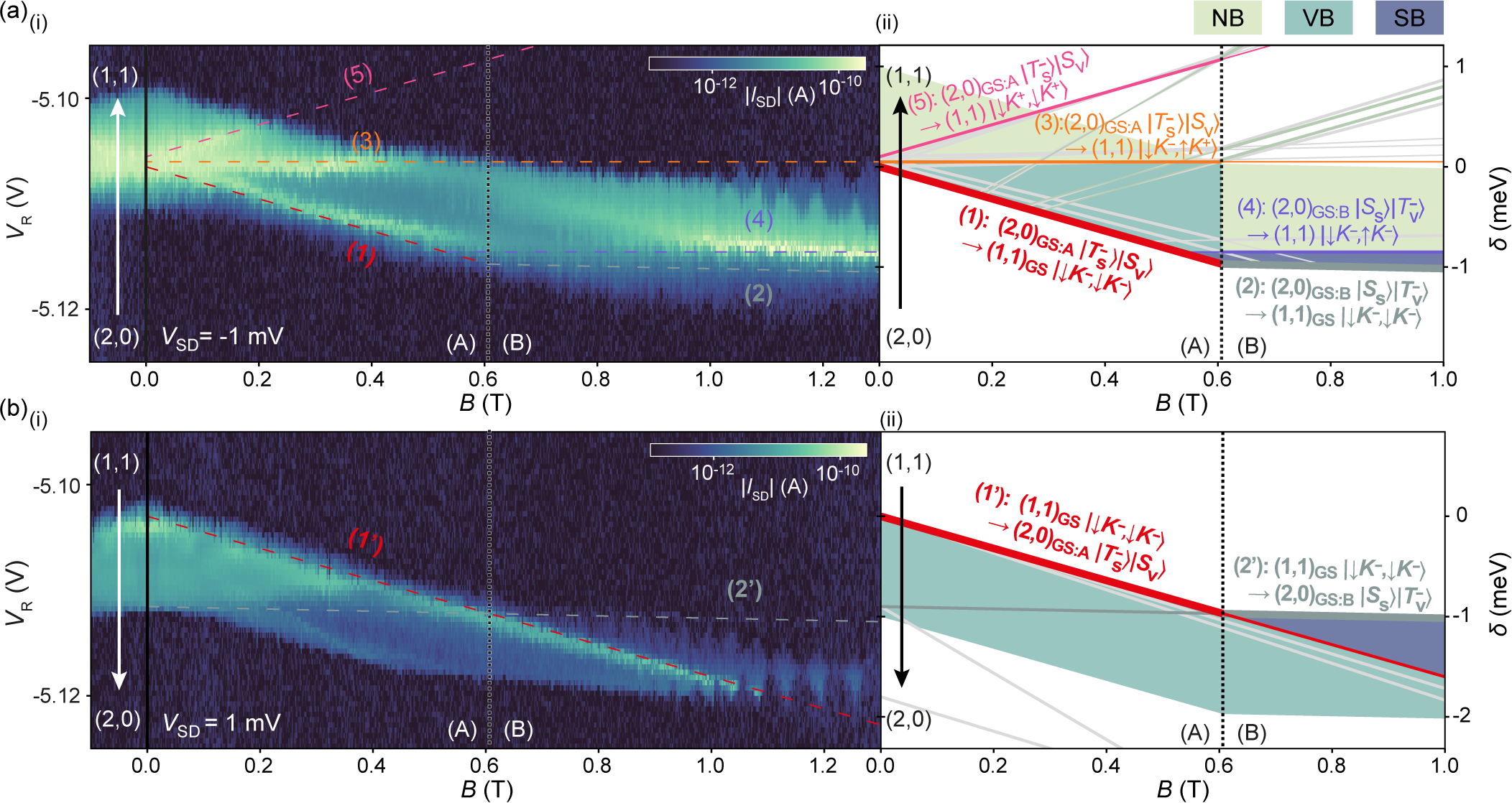}
	\caption{Evolution in magnetic field of GS transitions for (a) negative and (b) positive source--drain bias $V_\mathrm{SD}$. Evolution in magnetic field for (i) line-cut along the $\delta$-axis (dashed arrows in Fig.~\ref{figtriangle}A), and (ii) calculated transitions with $E_\mathrm{ex}=\SI{0.9}{\milli eV}$, $\Delta_\mathrm{SO}=\SI{80}{\micro eV}$, $g_\mathrm{v}=28$ and $g_\mathrm{s}=2$, from $(2,0)_\mathrm{GS}$ in (a,ii), and from $(1,1)_\mathrm{GS}$ in (b,ii). Except for the GS--GS transition $(2)$ and $(2')$, transitions requiring spin flips (sketched in light gray in (ii)) are not labeled. The thicker lines in (ii) represent the GS--GS transitions, and hence define the baselines of the bias triangles. Yellow, blue, and purple represent non-blockaded (NB), valley-blockaded (VB), and spin-blockaded (SB) regions, respectively. At high field, oscillations periodic in $1/B$ are observed for both bias directions \cite{LucaSdH}.}
	\label{figlevels}
\end{figure*}

A finite bias voltage $V_\mathrm{SD}$ expands triple points into finite-bias triangles, whose orientation depends on the sign of $V_\mathrm{SD}$. Within the bias triangles, energies of the relevant states are tuned into the bias window, such that these states are accessible and allow for charge transport. We look at the two-electron transport at the triple points encircled in Fig.~\ref{figstart}(b), where transitions between the two-electron charge occupation $(1,1)$ (one electron on each dot) and $(2,0)$ (both electron on the left dot) states govern the transport. (Note that we observe the same physics around the charge occupation $(1,1)$ and $(0,2)$, only swapping the role of the left and the right dot, see Supplemental Materials S6.) An example of current map of bias-triangles at $B=\SI{800}{\milli T}$ is shown in Fig.~\ref{figstart}(c). In one bias direction [Fig.~\ref{figstart}(c,i)], the high conductance bias triangles are complete; in the other bias direction [Fig.~\ref{figstart}(c,ii)] however, the triangles appear smaller with a missing baseline. Here, energetically allowed transitions are blocked by the Pauli exclusion principle, blocking charge transport and suppressing current, demonstrating the Pauli blockade effect. Comparing line-cuts [Fig.~\ref{figstart}(c,iii)] along the dashed arrows we see: At the baselines (black dots), for negative bias the peak current is $\sim\SI{180}{\pico A}$, whereas for positive bias it is masked by the noise floor ($\sim\SI{300}{\femto A}$). In the remaining part of the positive-bias triangles, the current is $\sim 3$ times weaker compared to that in the negative-bias ones. In the following discussion, and with the aid of schematics [Fig.~\ref{figstart}(c,iv and v)], we will attribute the missing baseline to spin blockade, and the weaker current in the tip of the triangle to valley blockade.

A thorough understanding of the relevant $(1,1)$ and $(2,0)$ states is crucial for interpreting the nature of the blockade. Here, we introduce the recently established level spectrum of one- and two-particle states in single BLG QDs \cite{annikaexcitedstates, AnnikaKondo,Luca1001,Lucacrossover,RebekkaPRL,Tunabledot,AngelikaQuartetStates,Angelika2021}, and limit our discussion to the lowest orbital state (the next orbital state is $>\SI{1.7}{\milli eV}$ higher in energy, see Supplemental Materials S2 and S3). 

Within the first energy shell, the single-dot--single-particle states are four-fold degenerate. A small spin--orbit coupling $\Delta_\mathrm{SO}\sim\SI{80}{\micro eV}$ \cite{AnnikaKondo,Luca1001} splits them into two Kramer pairs, $\ket{\downarrow K^-}$ and $\ket{\uparrow K^+}$, and $\ket{\uparrow K^-}$ and $\ket{\downarrow K^+}$ (see Supplemental Materials S1). A magnetic field splits the spin states by $\Delta E_\mathrm{\uparrow/\downarrow}=\pm g_\mathrm{s}\mu_\mathrm{B}B/2$ in energy, where $g_\mathrm{s}=2$ \cite{MariusPRX, RebekkaPRL, annikaexcitedstates}. Analogously, the valleys $K^{\pm}$ couple to a perpendicular magnetic field with $\Delta E_\mathrm{K^{\pm}}=\pm g_\mathrm{v}\mu_\mathrm{B}B_\perp/2$, linearly in the low-field limit. The valley g-factor $g_\mathrm{v}$ is an order of magnitude larger than the spin g-factor $g_\mathrm{s}$ \cite{MariusPRX,annikaexcitedstates,Tunabledot}.

For weakly coupled double dots, exchange interaction is negligible between two electrons, each residing on one dot. The $(1,1)$ basis states can therefore be approximated as product states of two sets of single-dot--single-particle states, generating 16 $(1,1)$ basis states with 10 distinct energies $E_\mathrm{(1,1)}$ that are sums of the energies of the two sets (shown in Supplemental Materials S4). The ground state $(1,1)_\mathrm{GS}$ is always $\ket{\downarrow K^-,\downarrow K^-}$.

Contrarily, the $(2,0)$ states are better described in the basis of spin and valley singlet and triplets. We consider only the lower symmetric orbital wave function, so that the product of the valley and spin states is necessarily anti-symmetric. In this way, we reduce the 16 basis states to 6. Evolution of this single-dot two-particle spectrum \cite{AngelikaQuartetStates,annikaexcitedstates,Angelika2021} in a perpendicular magnetic field is sketched in Fig.~\ref{figstart}(d). At low field, the spin-triplet--valley-singlet states, $\ket{T^{-/0/+}_\mathrm{s}}\ket{S_\mathrm{v}}$, are lower in energy, than the spin-singlet--valley-triplets, $\ket{S_\mathrm{s}}\ket{T^{-/+}_\mathrm{v}}$, by the exchange energy $E_\mathrm{ex}$. At high field, $\ket{S_\mathrm{s}}\ket{T^-_\mathrm{v}}$ is lowered enough in energy due to coupling with the magnetic field to become the GS. We therefore separate the discussion into two regimes, where the $(2,0)_\mathrm{GS}$ is $\ket{T^-_\mathrm{s}}\ket{S_\mathrm{v}}$ in regime A, and $\ket{S_\mathrm{s}}\ket{T^-_\mathrm{v}}$ in regime B.

With the knowledge of the expected $(1,1)$ and $(2,0)$ states, we look in Fig.~\ref{figtriangle} at finite-bias triangles with low inter-dot coupling, where occurrences of elastic tunnelings are observed as current resonances in the left and right dot energy difference, i.e. the detuning axis $\delta=\epsilon_\mathrm{L}-\epsilon_\mathrm{R}$. The different appearances of the three pair of triangles indicate distinct sets of transitions involved, but the current at positive bias is always suppressed compared to negative bias: Unless higher orbital states are within reach in the bias window, due to the Pauli exclusion principle, there exist no $(2,0)$ states matching both spin and valley quantum numbers of the $(1,1)_\mathrm{GS}$ $\ket{\downarrow K^-,\downarrow K^-}$.

In regime A (Fig.~\ref{figtriangle}A), states involved in the GS--GS transitions $(1),(1')$ (red) have mismatching valleys. Unless another transition channel from the GSs exists, allowing the electron to by-pass this blockade, an electron loaded into the GSs would be stuck and would suppress the current at the baselines, until a valley-flip event occurs. 

However, at $B=\SI{0}{T}$ the baseline for $(2,0)\rightarrow(1,1)$ [Fig.~\ref{figtriangle}A(a,i)] is strong, with a peak current of $\sim\SI{350}{\pico A}$ [Fig.~\ref{figtriangle}A(a,iii)], as the valley-blockaded transition $(1)$ (red) is easily circumvented by the non-blockaded transition $(3)$ (orange), because the $(1,1)$ states $\ket{\downarrow K^-,\downarrow K^-}$ and $\ket{\downarrow K^-,\downarrow K^+}$ (or $\ket{\downarrow K^+,\downarrow K^-}$) are nearly degenerate in energy at $B=\SI{0}{T}$.
At $B=\SI{200}{\milli T}$ [Fig.~\ref{figtriangle}A(b)], these $(1,1)$ states are valley split in energy by $g_\mathrm{v}\mu_\mathrm{B}B$. Hence, transitions $(1)$ and $(3)$ no longer occur at the same energy, as $(1)$ is lowered in detuning while $(3)$ stayed the same. We see therefore in Fig.~\ref{figtriangle}A(b,i) a valley-blockaded region with suppressed current, which at higher $\delta$ is lifted by the onset of transition $(3)$.

By contrast, for $(1,1)\rightarrow(2,0)$ [Fig.~\ref{figtriangle}A(a and b, ii)] the valley blockade cannot be circumvented by another transition. Current at the baseline is suppressed by a factor larger than five at $B=\SI{0}{T}$, and larger than ten at $B=\SI{200}{\milli T}$. This is because the next available $(2,0)$ state accessible from the $(1,1)_\mathrm{GS}:\ket{\downarrow K^-,\downarrow K^-}$ with matching valleys is $\ket{S_\mathrm{s}}\ket{T^-_\mathrm{v}}$. Transition $(2')$ (gray) to this state is not only higher in energy, but also requires a spin flip. Even at finite $\delta$ where enough energy is provided, no lifting of the valley blockade via this spin-mismatched transition is observed. 

The valley blockade is lifted at $B=\SI{0}{T}$ at the outer edges of the triangles [Fig.~\ref{figtriangle}A(a,ii), yellow arrows], with current similar to the non-blockaded inelastic current. At the edges, the GS electrochemical potential of the right dot $\mu_\mathrm{R,(1,1)GS}$ is aligned with the drain $\mu_\mathrm{D}$ \cite{JohnsonBlockade,RMPspin, Thomasbook}, allowing an electron with blockaded quantum numbers to tunnel back into the lead, in exchange for one with quantum numbers that allow the transport to continue. This lifting is no longer observed at finite field $B=\SI{200}{\milli T}$ in Fig.~\ref{figtriangle}A(b,ii), as the blockaded $(1,1)_\mathrm{GS}$ $\ket{\downarrow K^-,\downarrow K^-}$ and the non-blockaded $(1,1)$ state $\ket{\downarrow K^-,\downarrow K^+}$ are split by $g_\mathrm{v}\mu_\mathrm{B}B$, more than the thermal energy. 

In regime B at higher field, when $(2,0)_\mathrm{GS}$ becomes $\ket{S_\mathrm{s}}\ket{T^-_\mathrm{v}}$ (Fig.~\ref{figtriangle}B), the GS--GS transitions $(2)$ and $(2')$ are spin-blockaded. However, for the $(2,0)\rightarrow(1,1)$ bias direction [Fig.~\ref{figtriangle}B(i)], the spin-blockaded transition $(2)$ (gray) can be circumvented via transition $(4)$ (purple) that is very close in energy (only a Zeeman splitting higher in detuning, see Supplemental Materials S5 for more details), with a peak current of $\SI{200}{\pico A}$ [Fig.~\ref{figtriangle}B(iii)].

For $(1,1)\rightarrow(2,0)$ [Fig.~\ref{figtriangle}B(ii)], the spin-blockade leakage current is smaller than the noise floor. The next available transition in detuning is the valley-blockaded transition $(1')$ (red) discussed above. This transition is observed at larger $\delta$, with a peak current of $\SI{10}{\pico A}$. Spin conservation during inter-dot tunneling is a stronger condition than valley conservation, as the valley blockaded transition $(1')$ lifts the spin blockade [Fig.~\ref{figtriangle}B(ii)], but the spin-blockaded transition $(2')$ cannot lift the valley blockade [Fig.~\ref{figtriangle}A(a and b, ii)]. When increasing inter-dot coupling, we enhance current from transport via non-elastic tunneling, and arrive at the Fig.~\ref{figstart}(c) shown before.

We inspect the line-cuts along the $\delta$-axis (dashed arrows in Fig.~\ref{figtriangle}A) in magnetic field for the continuous evolution of the identified transitions. The results are displayed in Fig.~\ref{figlevels}(a and b,i) for $(2,0)\rightarrow(1,1)$ and $(1,1)\rightarrow(2,0)$, respectively. The corresponding calculated transition energies are plotted in Fig.~\ref{figlevels}(a and b,ii) (see Supplemental Materials S4 for evolution of the states). $V_\mathrm{SD}$ opens up a bias window of $\SI{1}{\milli eV}$ starting from the baseline, shown as higher conductance regions in the measurements (a and b,i), and colored regions in the calculated transitions (a and b,ii). Beyond the bias window, electron occupancy is Coulomb blockaded in either $(1,1)$ or $(2,0)$.

For $(2,0)\rightarrow(1,1)$, in regime A, transitions $(1)$, $(3)$, and $(5)$ split linearly in energy with the magnetic field, with $g_\mathrm{v}\approx 28$, corresponding to the $(1,1)$ valley configurations $\ket{K^-,K^-}$, $\ket{K^-,K^+}$ ($\ket{K^+,K^-}$), and $\ket{K^+,K^+}$. The kink in the baseline [red and gray in Fig.~\ref{figlevels}(a)] indicates the change of GS--GS transitions from $(1)$ to $(2)$, caused by the change of the $(2,0)_\mathrm{GS}$ from $\ket{T^-_\mathrm{s}}\ket{S_\mathrm{v}}$ to $\ket{S_\mathrm{s}}\ket{T^-_\mathrm{v}}$.

For $(1,1)\rightarrow(2,0)$, only the valley-blockaded transition $(1')$ is observed. At high field, the bias window diminishes and the edges appear no longer parallel to $(1')$, but to $(2')$ instead. This indicates the change of the GS--GS transition from $(1')$ [red in Fig.~\ref{figlevels}(b)] to the spin-blockaded transition $(2')$ (gray), with resonance masked by the noise floor.


Regions of no-blockade, valley-blockade, and spin-blockade are labeled in Fig.~\ref{figlevels}(a and b,ii). Current strength in these regions decreases due to the blockade effect in this order [Fig.~\ref{figlevels}(a and b,i)]. The singlet--triplet energy splitting, crucial for spin-qubit operation, can be tuned in magnitude by magnetic field, or by tuning the valley g-factor with gate voltages \cite{Tunabledot}.

In conclusion, in our BLG QDs we show controlled switching between two regimes: At low perpendicular magnetic field, the (2,0) ground state is a spin-triplet-valley-singlet, allowing for observation of valley blockade; whereas at higher field, the spin-singlet-valley-triplet (2,0) ground state allows for the observation of robust spin blockade. These results demonstrate exquisite control over spin and valley states, thorough understanding of the intricate two-particle Hilbert space, and high sample quality of our BLG QDs. The observation of blockade paves the way for future graphene based spin and valley qubits. 

\nocite{supp}

\nocite{wang2013drytransferedge}

\section*{acknowledgments}

We thank P. Märki and T. Bähler as well as the FIRST staff for their technical support. We acknowledge funding from the Core3 European Graphene Flagship Project, the Swiss National Science Foundation via NCCR Quantum Science and Technology, and the EU Spin-Nano RTN network. R. Garreis acknowledges funding from the European Union’s Horizon 2020 research and innovation programme under the Marie Skłodowska-Curie Grant Agreement No. 766025. K.W. and T.T. acknowledge support from the Elemental Strategy Initiative conducted by the MEXT, Japan , Grant Number JPMXP0112101001, JSPS KAKENHI Grant Number 19H05790 and JP20H00354.

\section*{Data availability}
The data supporting the findings of this study is made available via the ETH Research Collection: Data repository: Pauli blockade of tunable two-electron spin and valley states in graphene quantum dots (2022), 10.3929/ethz-b-000528896.

\section*{Competing interests}
The authors declare no competing interests.
%

\newpage

\setcounter{section}{0} 

\renewcommand\thesection{S~\arabic{section}} 
\setcounter{figure}{0} 
\renewcommand\thefigure{S\arabic{figure}}

\setcounter{section}{0} 
\renewcommand\thesection{S~\arabic{section}} 

\setcounter{figure}{0}
\renewcommand\thefigure{S\arabic{figure}}

\section{S1.~Methods}

\label{section:methods}

\begin{figure}
	\includegraphics[width=8.5cm]{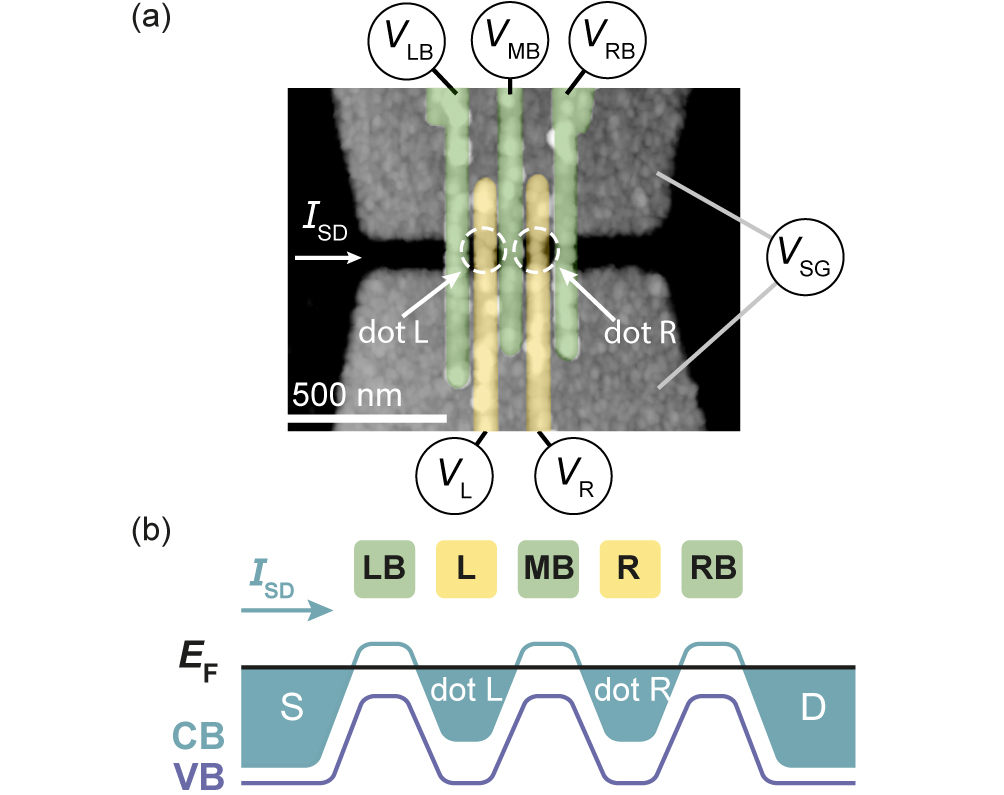}
	\caption{(a) False-colored AFM image of the sample used, where the colors of the gates corresponds to the ones shown in Fig.1(a). Voltages supplied to these gates are la belled. (b) Schematic sketch of the conduction band (CB) and valence band (VB) edge variation along the channel, underneath the respective gates.}
	\label{sampleS}
\end{figure}

The device is fabricated as described in \cite{Hiske2018electrostatically, MariusPRX}, and schematically depicted in Fig.~1(a). A false-colored AFM image of the sample is shown in Fig.~\ref{sampleS}(a). Stacked with the dry-transfer technique \cite{wang2013drytransferedge}, the van der Waals hetero-structure lies on a silicon chip with \SI{280}{nm} surface SiO$_2$. The stack consists of a bottom graphite back gate [dark gray in Fig.~1(a)], and on top of it a BLG flake (black) encapsulated in \SI{38}{nm} thick bottom and \SI{20}{nm} thick top hBN flakes (blue). Ohmic edge contacts (light grey) with Cr and Au of \SI{10} and \SI{60}{nm} thickness, respectively, are evaporated after etching through the top hBN flake with reactive ion etching. A pair of \SI{5}{nm} thick Cr, \SI{20}{nm} thick Au split gates (purple) are deposited on top, defining a $\SI{1}{\micro m}$ long, $\SI{100}{nm}$ wide channel. Separated by a layer of \SI{30}{nm} thick amorphous Al$_2$O$_3$ grown by atomic layer deposition, finger gates (yellow and green) of \SI{20}{nm} in width, and \SI{5}{nm} Cr and \SI{20}{nm} Au in thickness, lie across the channel. Neighboring finger gates are separated by $\SI{75}{nm}$ from center to center.

The spatial variation of the band-edge of the coupled double electron dot is sketched in Fig.~\ref{sampleS}(b). We utilize the tunable BLG band-gap $\Delta_\text{gap}$ that arises from the application of an electric displacement field perpendicular to the BLG sheet \cite{Ohta2006BLGband,mccann2006BLGband,Oostinga2008BLG}. Closely arranged finger gates allow for local control within the $\SI{100}{nm}$ wide channel formed by the split gates. Negative barrier gate voltages $V_\mathrm{LB,MB,RB}$ tune regions underneath into the gap, isolating our pair of coupled double electron dots from the $n$-type channel, providing individual control of the dot-lead \cite{mariusnanolett,Tunabledot}, and inter-dot tunnel coupling \cite{LucaInterdot}. Dots L, R are independently controlled by the plunger gate voltages $V_\mathrm{L,R}$. 

We perform measurements in He$^{3}$/ He$^4$ dilution refrigerators at electronic temperatures of around $\SI{150}{\milli K}$.

\section{S2.~One-electron states}

\label{section:1001}

\begin{figure}
	\includegraphics[width=8.5cm]{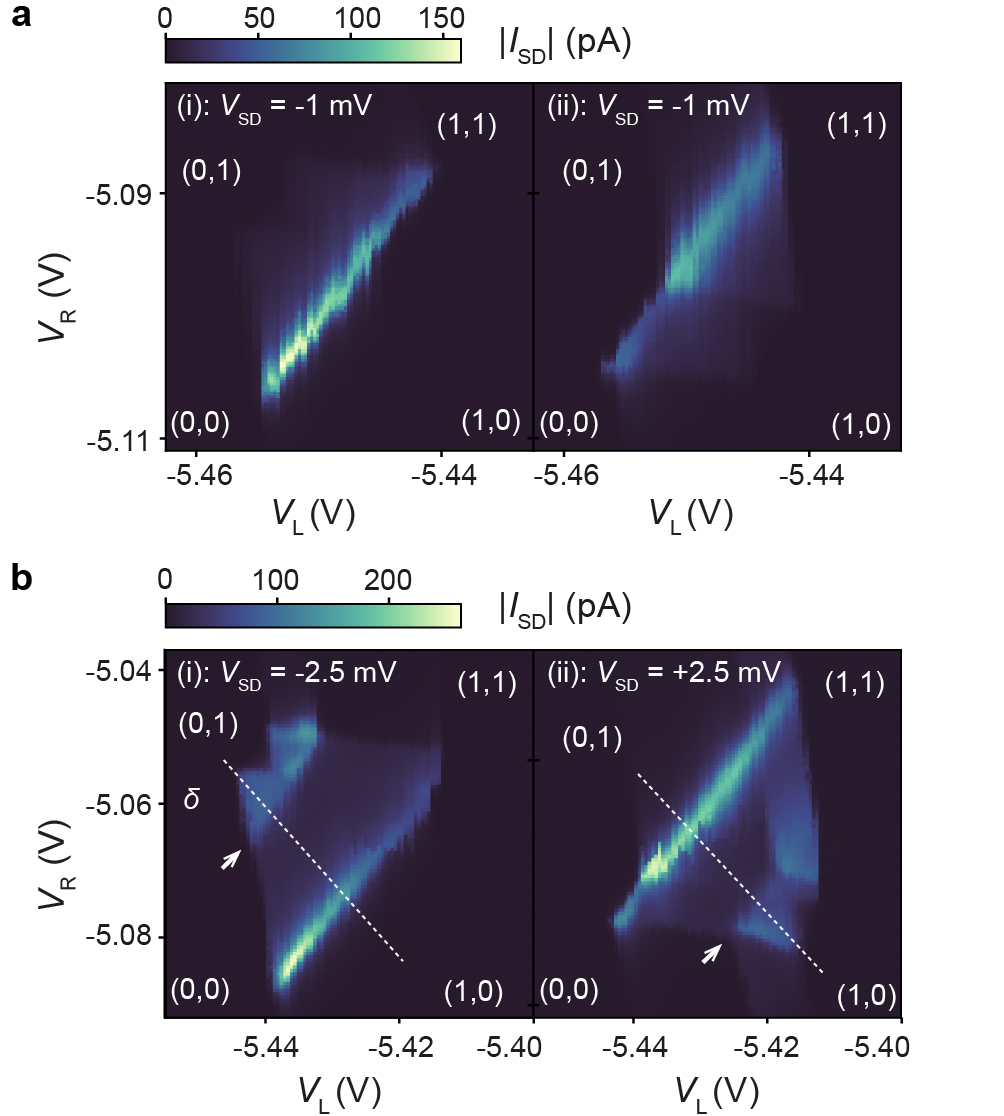}
	\caption{Finite-bias triangles near one-electron $(1,0)$ and $(0,1)$ states at $V_\mathrm{MB}\SI{-5.81}{V}$, $B=\SI{0}{T}$, for a $|V_\mathrm{SD}|=\SI{1}{\milli V}$ and b $|V_\mathrm{SD}|=\SI{2.5}{\milli V}$, for (i) negative (electron transport $(1,0)\rightarrow(0,1)$), and (ii) positive (electron transport $(0,1)\rightarrow(1,0)$) $V_\mathrm{SD}$. The conductance steps labeled by the white arrows in (b) at $|\delta|=\SI{1.7}{\milli eV}$ indicate access to the higher orbital state.}
	\label{1001triangle}
\end{figure}

\begin{figure*}
	\includegraphics[width=17.8cm]{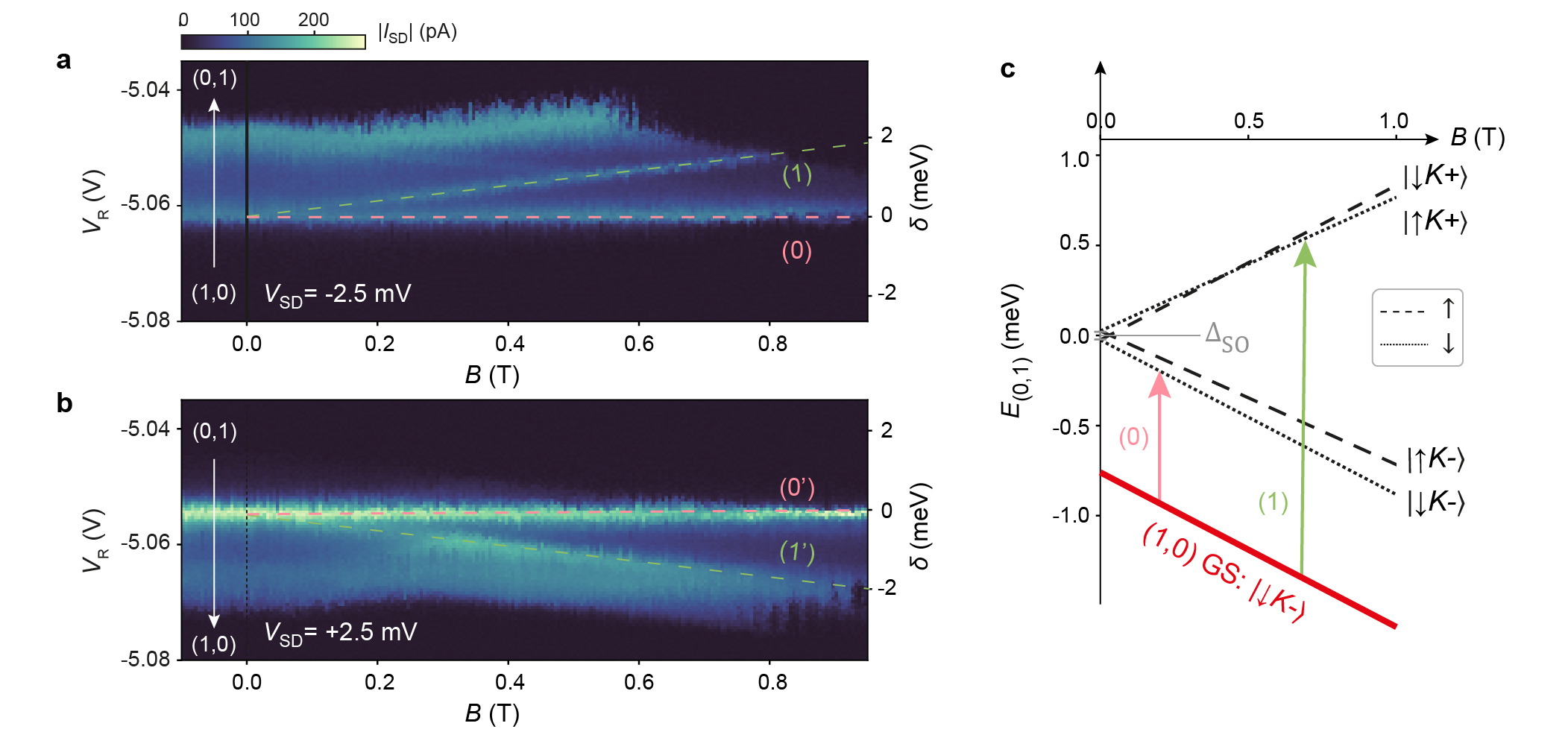}
	\caption{Evolution in magnetic field of GS transitions between one-electron states $(0,1)$ and $(1,0)$. Evolution of the line-cut along the $\delta$-axis (dashed lines) in Fig.~\ref{1001triangle}(b) for (a) negative ($V_\mathrm{SD}=\SI{-2.5}{\milli V}$, electron transport $(1,0)\rightarrow(0,1)$), and (b) positive ($V_\mathrm{SD}=\SI{2.5}{\milli V}$ electron transport $(0,1)\rightarrow(1,0)$) source--drain bias $V_\mathrm{SD}$. $V_\mathrm{SD}$ opens a bias window of $\SI{2.5}{\milli eV}$. Outside of the bias window, current is Coulomb blockaded at either $(0,1)$ or at $(1,0)$. (c) Calculated single-particle states $E_\mathrm{0,1}$ with $\Delta_\mathrm{SO}=\SI{80}{\micro eV}$, $g_\mathrm{v}=32$ and $g_\mathrm{s}=2$. GS of $(1,0)$ is sketched in red and shifted arbitrarily in the energy axis, where transitions from this  are labeled and identified in (a) as current peaks. At high field ($B>\SI{800}{\milli T}$), $(1,0)_{\mathrm{GS}}\ket{\downarrow K^-}$ oscillations periodic in $1/B$, i.e., Shubnikov--de Haas oscillations \cite{LucaSdH}.}
	\label{1001delta}
\end{figure*}

Behavior around the one-electron states $(1,0)$ and $(0,1)$ is studied. Finite bias triangle measurements are shown for two different source--drain bias sizes ($|V_\mathrm{SD}|=\SI{1}{\milli V},\SI{2.5}{\milli V}$) in Fig.~\ref{1001triangle}. In contrast to what was observed for the two-electron case Fig.~2(a), little asymmetry in bias-direction is observed here. The lack of blockade is expected for the one-electron states. An electron tunneling in from the source reservoir into the left dot (or from the drain into the right dot) has all spin and valley configurations at its disposal, since there is no electron in either of the dots before the transport as all states in both dots are unoccupied $(0,0)$. 

A larger source--drain bias [$|V_\mathrm{SD}|=\SI{2.5}{\milli V}$, Fig.~\ref{1001triangle}(b)] allows access to higher energy states. For both bias directions, a conductance step is observed at $|\delta|=\SI{1.7}{\milli eV}$, beyond which the single-dot orbital state higher in energy becomes accessible in the bias window, and transport via the higher orbital state can occur. 

We study the evolution in a perpendicular magnetic field for the line-cut along the detuning $\delta$-axis at constant total energy $\epsilon$, as indicated by the white dotted lines in Fig.~\ref{1001triangle}(b). The result at $|V_\mathrm{SD}|=\SI{2.5}{\milli V}$ is shown in Fig.~\ref{1001delta}(a) for negative (electron transport $(1,0)\rightarrow(0,1)$), and Fig.~\ref{1001delta}(b) for positive (electron transport $(0,1)\rightarrow(1,0)$) bias voltages. As expected the two magneto-spectroscopy maps appear similar in bias direction, each with two prominent resonance peaks, labeled transitions $(0)$ and $(0')$ (pink), and $(1)$ and $(1')$ (green). As shown in Fig.~\ref{1001delta}(c) (pink arrow), the GS--GS transition $(0)$ corresponds to an electron tunneling from one dot into the same state in the next dot, without spin or valley flip events. Whereas at higher $\delta$, when the magnetic-field-split higher valley states become accessible, the electron loaded into one dot has the additional option to tunnel into a state in the next dot with a different valley number (i.e., from $K^-$ to $K^+$), providing that a valley flip event occurs [transition $(1)$, green arrow in Fig.~\ref{1001delta}(c)]. This extra transport channel $(1)$ gives rise to a conductance peak due to elastic tunneling, and splits away from transition $(0)$ owning to the coupling between the valley states and the magnetic field. We extract a valley g-factor $g_\mathrm{v}=32$ from the slope of $(1)$ in detuning, agreeing with previous measurements \cite{Tunabledot,MariusPRX,annikaexcitedstates}. The small difference with the value from the two-electron case $g_\mathrm{v}=28$ can be attributed to the different plunger gate voltages \cite{Tunabledot}. The small spin--orbit coupling term $\Delta_\mathrm{SO}\sim\SI{80}{\micro V}$ \cite{Luca1001,AnnikaKondo} is not resolved in this measurement.

For the calculated energies and transitions here we have assumed the same valley g-factor $g_\mathrm{v}$ for both the left and the right dots. If $g_\mathrm{v}$ were to be different between the two dots, the GS--GS transition $(0)$ and $(0')$ would have a finite slope in magnetic field, corresponding to $|g_{\mathrm{v,L}}-g_{\mathrm{v,R}}|$. Here $(0)$ stays roughly constant in $\delta$ in magnetic field, indicating the symmetry of the two dots \cite{Tunabledot}.

If the spin-flip event were to be common, we would naturally also expect to see current resonances corresponding to transitions between states of opposite spins, e.g., from $\ket{\downarrow K^-}$ to $\ket{\uparrow K^-}$. The detuning of these resonances would evolve in the magnetic field with a slope corresponding to the spin g-factor $g_\mathrm{s}$. Such a feature is however missing from the measurement, indicating the strong conservation of spin upon inter-dot tunneling, a similar conclusion as the two-electron study. 

The study of these one-electron case measurements matches with the expectation from the single-particle levels \cite{AnnikaKondo, Luca1001}, and also with previous studies \cite{Luca1001}, which confirms our correct accounting of carrier numbers in the double dot system.

\section{S3.~Higher orbital state}
\label{section:2.5mV}
\begin{figure}
	\includegraphics[width=8.5cm]{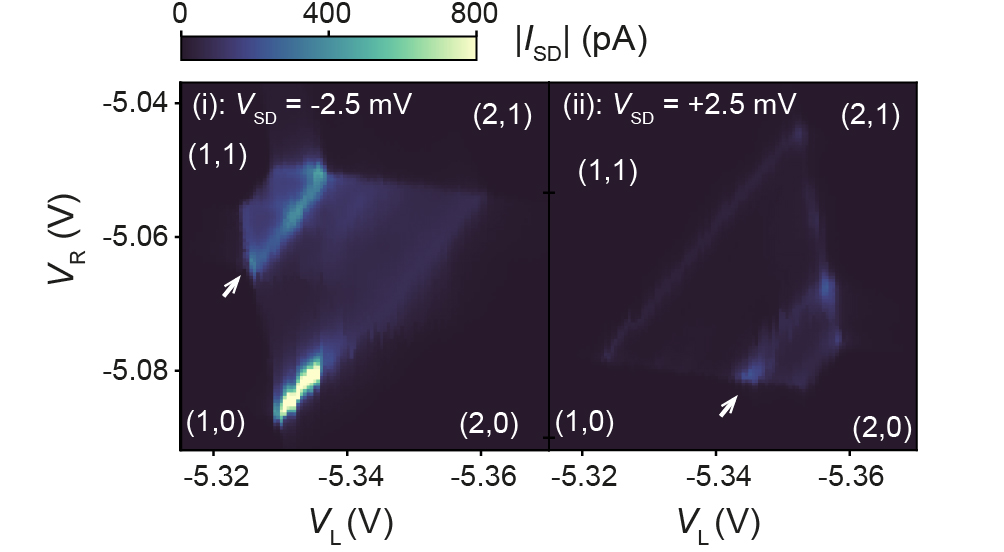}
	\caption{Low-field regime with spin-triplet--valley-singlet (2,0) ground state. Finite-bias triangles near the two-electron $(1,1)$ and $(2,0)$ states at $V_\mathrm{MB}\SI{-5.81}{V}$, $B=\SI{0}{T}$, for (i) negative ($V_\mathrm{SD}=\SI{-2.5}{\milli V}$, electron transport $(1,0)\rightarrow(0,1)$), and (ii) positive ($V_\mathrm{SD}=\SI{2.5}{\milli V}$ electron transport $(0,1)\rightarrow(1,0)$) source--drain bias $V_\mathrm{SD}$. The conductance steps labeled by the white arrows in (b) at $|\delta|=\SI{1.7}{\milli eV}$ indicate access to the higher orbital state, lifting the valley blockade in (ii).}
	\label{fig11202p5m}
\end{figure}

The two-electron bias triangles at $\SI{0}{T}$ at higher source--drain bias ($V_\mathrm{SD}=\SI{2.5}{\milli V}$) than that studied in the main text is shown in Fig.~\ref{fig11202p5m}. They are the same as the ones shown in Fig.~2, apart from the extra conductance step (indicated by the white arrows) at $|\delta|=\SI{1.7}{\milli eV}$, accessible now with the larger bias window. This conductance step occurs at the same detuning as that for the one-electron case shown in Fig.~\ref{1001triangle}, and corresponds to the higher orbital state.

\section{S4.~Two-electron state energies and transitions}
\label{section:energies}

\begin{figure*}
	\includegraphics[width=17.8cm]{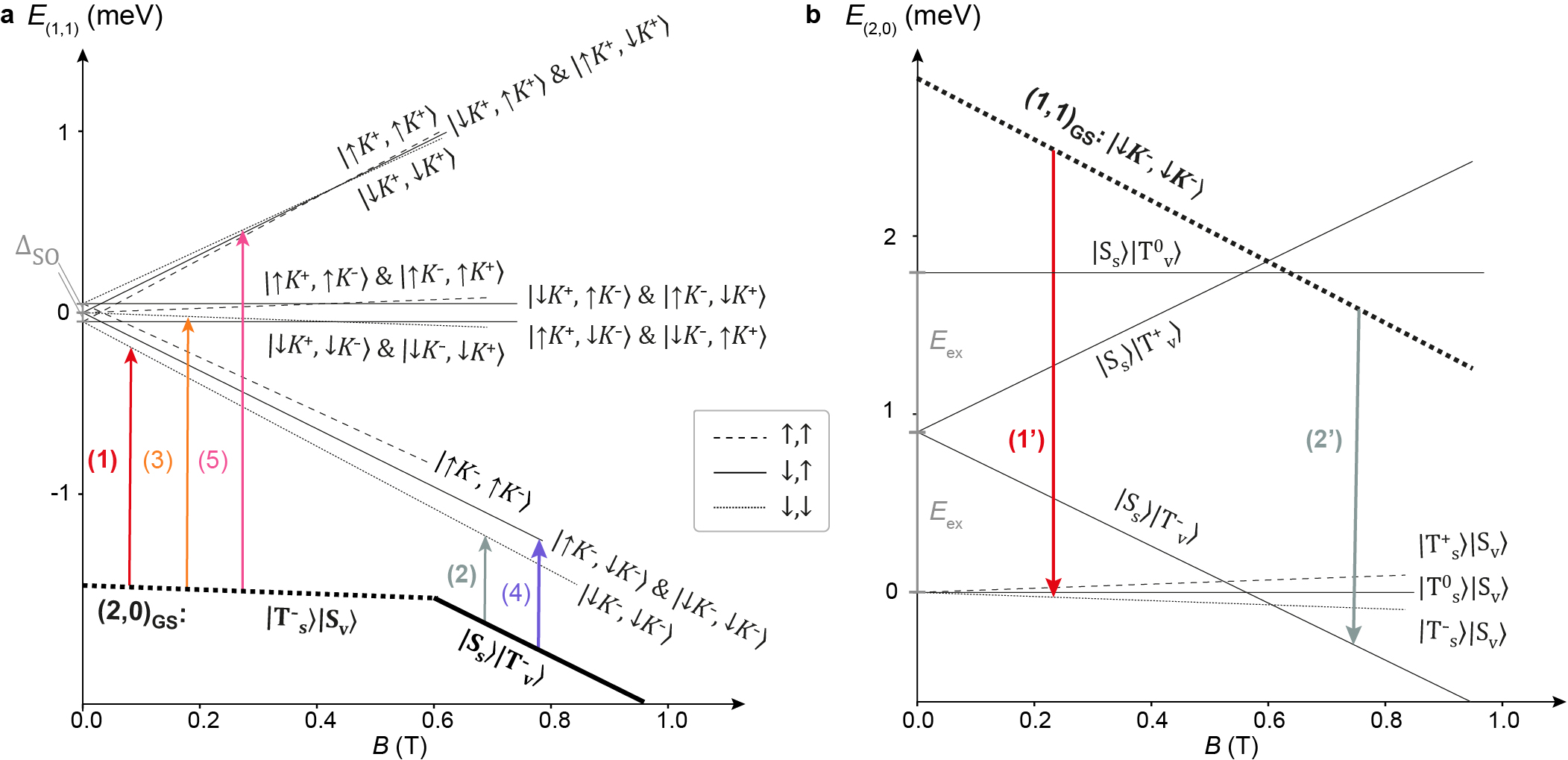}
	\caption{Calculated evolution in magnetic field of the two-electron energies, for (a) $E_\mathrm{(1,1)}$, when two electrons each reside on one dot, and for (b) $E_\mathrm{(2,0)}$ when both electrons reside on the same dot, with $E_\mathrm{ex}=\SI{0.9}{\milli eV}$, $\Delta_\mathrm{SO}=\SI{80}{\micro eV}$, $g_\mathrm{v}=28$, and $g_\mathrm{s}=2$. The $(2,0)$ GSs in (a), and the $(1,1)$ GSs in (b) are sketched in bold lines and shifted arbitrarily in energy for clarity. The transitions that originate from the ground states are labeled as discussed in the main text.}
	\label{figenergies}
\end{figure*}

The calculated evolution in magnetic field of the two-electron state energies are shown in Fig.~\ref{figenergies}, with the relevant GS transitions labeled. All non-spin flipped transitions originating from the $(2,0)_\mathrm{GS}$ in Fig.~\ref{figenergies}(a), and originating from the $(1,1)_\mathrm{GS}$ in Fig.~\ref{figenergies}(b), are labeled. 

\section{S5.~Blockade in negative bias directions}

In the main manuscript, we have discussed extensively about the spin blockade occurring in the positive bias direction. However, there could also be spin blockade occurring for the negative bias direction, at high magnetic field in regime B. At regime B, the $(1,1)$ ground state is both spin- and valley- polarized. However, the $(2,0)$ ground state in regime B, $(2,0)\ket{S_\mathrm{s}}\ket{T^-_\mathrm{v}}$, is only valley polarized, but spin-mixed. Therefore, transition from this $(2,0)$ ground state to the $(1,1)$ ground state (labeled transition (2) in Fig.2 and Fig.3 in the main text, and in Fig.S5) is blocked by the spin selection rule, hence demonstrates spin blockade.

At $B=\SI{800}{mT}$ where we did our measurement however, the spin Zeeman splitting is very small ($~\SI{90}{\mu eV}$). One can, look at the size of the negative bias triangles shown in Fig.2 in the main text: the triangles at $B=\SI{800}{mT}$ [Fig.2B(i)] is slightly smaller than that at $B=\SI{200}{mT}$ [Fig.2A(b,i)]. This effect is however not prominent in the resolution of the measurement.

Regions corresponding to this spin blockade is also labeled in the schematics in Fig.3(a,ii) of the main text. The measurement in Fig.3(a,i) however cannot resolve this region.

Blockade in the negative bias direction is hard to confirm, when the bias triangles in the positive direction also shows blockade. In this case, baselines of triangles in both bias directions are missing and suppressed by the spin-blockade effect, and one can therefore not identify blockade of the larger triangle in the negative bias direction as we cannot compare its size with 'less-blockaded' triangles. 

\begin{figure*}
	\includegraphics[width=17.8cm]{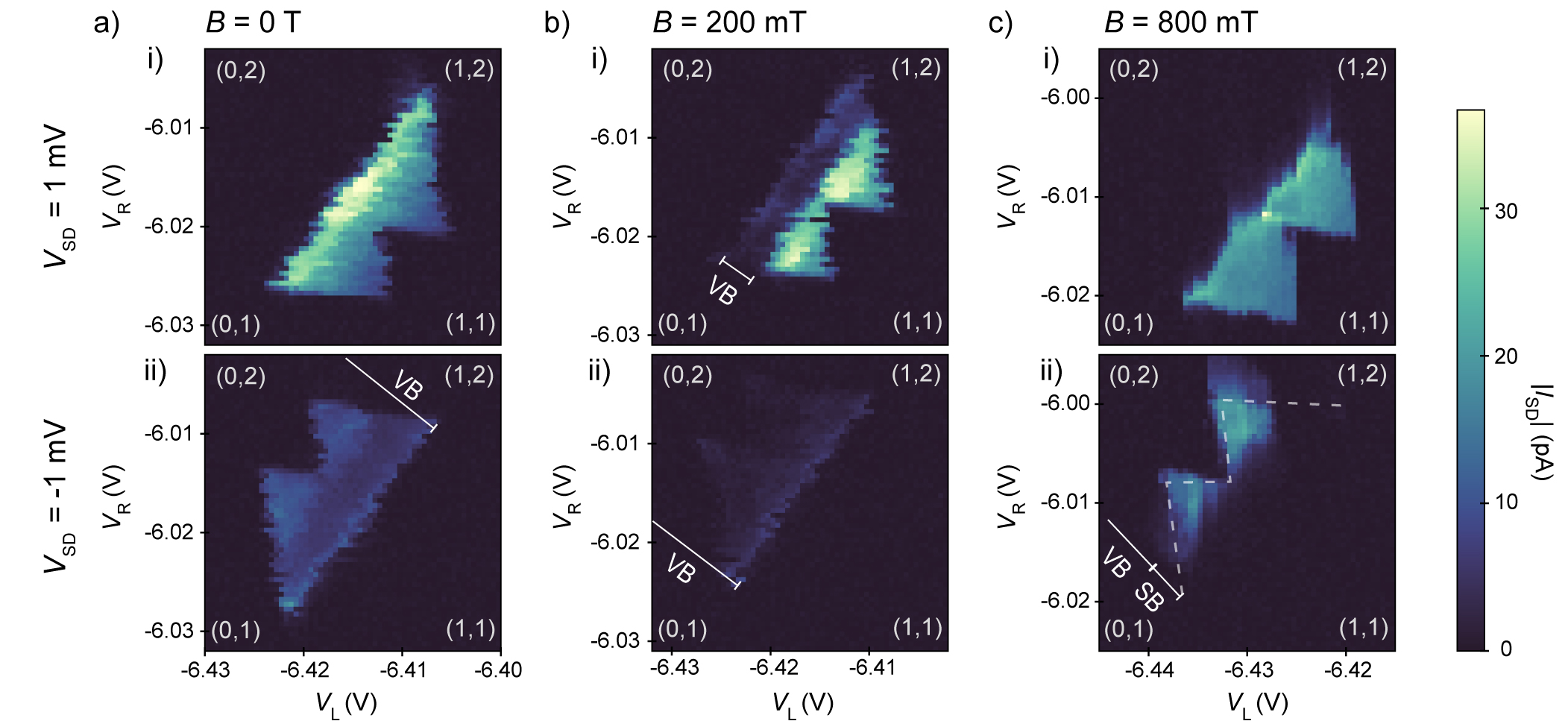}
	\caption{Finite-bias triangles at $V_\mathrm{BG}=\SI{5.0}{V}$, $V_
	\mathrm{MB}=\SI{-6.22}{V}$ at in regime A with $(0,2)_\mathrm{GS}:\ket{T^-_\mathrm{s}}\ket{S_\mathrm{v}}$ at $B=$ (a) $\SI{0}{T}$ and (b) $\SI{200}{mT}$, and in regime B with $(0,2)_\mathrm{GS}:\ket{S_\mathrm{s}}\ket{T^-_\mathrm{v}}$ at (c) $B=\SI{800}{mT}$ for (i) positive (electron transport $(0,2)\rightarrow(1,1)$), and (ii) negative (electron transport $(1,1)\rightarrow(0,2)$) source--drain bias $V_\mathrm{SD}$.}
	\label{fig2011triangles}
\end{figure*}

\section{S6.~Blockade in the opposite charge occupation}
In the main text of the manuscript we focused on the blockade between $(1,1)$ and $(2,0)$ charge occupation, i.e. between each electron on one dot, and both electrons on the left dot. Similarly we can also observe Pauli blockade in the opposite charge arrangement, between $(1,1)$ each electron on one dot, and $(0,2)$ both electron on the right dot. Here the physics is exactly the same, just that the role of left and right dot is exchanged. Fig.~\ref{fig2011triangles} show bias triangles near the $(1,1)$ and $(0,2)$ charge states at $B=\SI{0}{T},\SI{200}{mT}$ and $\SI{800}{mT}$, for both bias directions. The interpretation of Pauli blockade here is exactly the same as that discussed in Fig.2 in the main text.

\end{document}